\pdfoutput=1

\documentclass[final,5p,times,twocolumn]{elsarticle}

\usepackage{amsmath,amssymb,amsthm}
\usepackage[boxed]{algorithm}
\usepackage[tight,footnotesize]{subfigure}

\graphicspath{{.}{pics/}{pics/new/}}

\begin{document}

\begin{frontmatter}



\title{Variational Bayesian Adaptation of Noise Covariances in
Non-Linear Kalman Filtering}


\author{Simo~S\"arkk\"a and Jouni Hartikainen}

\address{Aalto University, Finland}

\begin{abstract}
  This paper is considered with joint estimation of state and
  time-varying noise covariance matrices in non-linear stochastic
  state space models. We present a variational Bayes and Gaussian
  filtering based algorithm for efficient computation of the
  approximate filtering posterior distributions. The Gaussian
  filtering based formulation of the non-linear state space model
  computation allows usage of efficient Gaussian integration methods
  such as unscented transform, cubature integration and Gauss-Hermite
  integration along with the classical Taylor series approximations.
  The performance of the algorithm is illustrated in a simulated
  application.
\end{abstract}


\end{frontmatter}

\section{Introduction}
In this paper, we consider Bayesian inference on the state $x_k$ and
noise covariances $\Sigma_k$ in heteroscedastic non-linear stochastic
state space models of the form
\begin{equation}
\begin{split}
  x_k &\sim \mathrm{N}\left(f(x_{k-1}),Q_k\right) \\
  y_k &\sim \mathrm{N}\left(h(x_k),\Sigma_k\right) \\
  \Sigma_k &\sim p(\Sigma_k\,|\,\Sigma_{k-1}),
\end{split}
\label{eq:ssmodel1}
\end{equation}
where $x_k \in \mathbb{R}^n$ is the state at time step $k$, and $y_k
\in \mathbb{R}^d$ is the measurement, $Q_k$ is the known process noise
covariance and $\Sigma_k$ is the measurement noise covariance.
The non-linear functions $f(\cdot)$ and $h(\cdot)$ form the dynamic
and measurement models, respectively, and the last equation defines
the Markovian dynamic model for the dynamics of the unknown noise
covariances $\Sigma_k$. 

The purpose is to estimate the joint posterior (filtering) distribution
of the states and noise covariances:
\begin{equation}
\begin{split}
  p(x_k,\Sigma_k\,|\,y_{1:k}),
\end{split}
\end{equation}
where we have introduced the notation $y_{1:k} = y_1,\ldots,y_k$.

If the parameters $Q_k$ and $\Sigma_k$ in the model
\eqref{eq:ssmodel1} were known, the state estimation problem would
reduce to the classical non-linear (Gaussian) filtering problem
\cite{Jazwinski:1970,Maybeck:1982}. However, here we consider the
case, where the noise covariances $\Sigma_k$ are unknown.  The formal
Bayesian filtering solution for general probabilistic state space
models, including the one considered here, is well known (see, e.g.,
\cite{Jazwinski:1970}) and consist of the Chapman-Kolmogorov equation
on the prediction step and Bayes' rule on the update step. However,
the formal solution is computationally intractable and we can only
approximate it.

In a recent article, S\"arkk\"a and Nummenmaa
\cite{Sarkka+Nummenmaa:2009} introduced the variational Bayesian
adaptive Kalman filter (VB-AKF), which can be used for estimating the
measurement noise variances along with the state in linear state space
models. In this paper, we extend the method to allow estimation of the
full noise covariance matrix and non-linear state space models.  The
idea is similar to what was recently used by Pich\'e et al.
\cite{Piche+Sarkka+Hartikainen:2012} in the context of oulier-robust
filtering, which in turn is based on the linear results of
\cite{Agamennoni+Nieto+Nebot:2011}.



We use the Bayesian approach and use the free form variational
Bayesian (VB) approximation (see, e.g.,
\cite{Jaakkola:2001,Beal:2003,Lappalainen+Miskin:2000}) for the joint
filtering posteriors of states and covariances, and the Gaussian
filtering approach \cite{Ito+Xiong:2000,Wu+Hu+Wu+Hu:2006} for handling
non-linear models.  The Gaussian filtering approach allows us also to
utilize more general methods such as unscented Kalman filter (UKF)
\cite{Julier+Uhlmann+Durrant-Whyte:2000}, Gauss-Hermite Kalman filter
(GHKF) \cite{Ito+Xiong:2000}, cubature Kalman filter (CKF)
\cite{Arasaratnam+Haykin:2009}, and various others
\cite{Norgaard+Poulsen+Ravn:2000,Lefebvre+Bruyninckx+DeSchutter:2002,Deisenroth+Huber+Hanebeck:2009}
along with the classical methods \cite{Jazwinski:1970,Maybeck:1982}.

The variational Bayesian approach has been applied to parameter
identification in state space models also in
\cite{Smidl+Quinn:2006,Beal+Ghahramani:2001,Valpola+Harva+Karhunen:2004}
and other Bayesian approaches are, for example, Monte Carlo methods
\cite{Storvik:2002,Djuric+Miquez:2002,Sarkka:2006c} and multiple model
methods \cite{Li+Bar-Shalom:1994}. It is also possible to do adaptive
filtering by simply augmenting the noise parameters as state
components \cite{Maybeck:1982} and use, for example, above-mentioned
Gaussian filters for estimation of the state and parameters.

\subsection{Gaussian Filtering}
If the covariances in the model \eqref{eq:ssmodel1} were known, the
filtering problem would reduce to the classical non-linear (Gaussian)
optimal filtering problem \cite{Jazwinski:1970,Maybeck:1982}. This
non-linear filtering problem can be solved in various ways, but one
quite general approach is the Gaussian filtering
approach \cite{Maybeck:1982,Ito+Xiong:2000,Wu+Hu+Wu+Hu:2006}, where
the idea is to assume that the filtering distribution is approximately
Gaussian. That is, we assume that there exist means $m_k$ and
covariances $P_k$ such that
\begin{equation}
  p(x_{k}~|~y_{1:k}) \approx \mathrm{N}(x_{k}~|~m_k,P_k).
\end{equation}
The Gaussian filter prediction and update
steps can be written as follows \cite{Ito+Xiong:2000}:
  \begin{itemize}
  \item Prediction:
  \begin{equation}
  \begin{split}
    m^-_{k} &=
      \int f(x_{k-1}) \,
        \mathrm{N}(x_{k-1}\,|\,m_{k-1},P_{k-1}) \, \mathrm{d} x_{k-1} \\
    P^-_{k} &= \int (f(x_{k-1}) - m^-_{k}) \,
       (f(x_{k-1}) - m^-_{k})^T \\
       &\qquad \times
       \mathrm{N}(x_{k-1}\,|\,m_{k-1},P_{k-1}) \, \mathrm{d} x_{k-1} 
       + Q_{k}.
  \end{split}
  \label{eq:adf_predict1}
  \end{equation}

  \item Update:
  \begin{equation}
  \begin{split}    
    \mu_{k} &= 
       \int h(x_{k}) \,
        \mathrm{N}(x_{k}\,|\,m^-_{k},P^-_{k}) \, \mathrm{d} x_k \\
    S_{k} &= \int (h(x_{k}) - \mu_{k}) \,
       (h(x_{k}) - \mu_{k})^T \\
     &\qquad \times 
       \mathrm{N}(x_{k}\,|\,m^-_{k},P^-_{k}) \, \mathrm{d} x_{k}
       + \Sigma_{k} \\
    C_k &=  \int (x_k - m^-) \,
                  (h(x_k) - \mu_k)^T \,
                  \mathrm{N}(x_k\,|\,m^-_k,P^-_k) \, \mathrm{d} x_k \\
    K_{k} &= C_k \, S^{-1}_{k} \\
    m_{k} &= m^-_{k} + K_{k} \, (y_k - \mu_k) \\
    P_{k} &= P^-_{k} - K_{k} \, S_{k} \, K^T_{k}.
  \end{split}
  \label{eq:adf_update1}
  \end{equation}
  \end{itemize}
With different selections for the Gaussian integral approximations,
we get different filtering algorithms \cite{Wu+Hu+Wu+Hu:2006}.

\subsection{Variational Approximation}
In this paper, we approximate the joint filtering distribution of
the state and covariance matrix with the free-form variational
Bayesian (VB) approximation (see, \emph{e.g.},
\cite{Jaakkola:2001,Beal:2003,Lappalainen+Miskin:2000,Smidl+Quinn:2006}):
\begin{equation}
  p(x_k,\Sigma_k\,\vert\,y_{1:k})
  \approx Q_x(x_k) \, Q_{\Sigma}(\Sigma_k),
\label{eq:vbapp}
\end{equation}
where $Q_x(x_k)$ and $Q_{\Sigma}(\Sigma_k)$ are the yet unknown
approximating densities. The VB approximation can be formed by
minimizing the Kullback-Leibler (KL) divergence between the true
distribution and the approximation:
\begin{equation}
\begin{split}
&\textrm{KL}[Q_x(x_k) \,
  Q_{\Sigma}(\Sigma_k)\,||\,p(x_k,\Sigma_k\,|\,y_{1:k})] \\ &= 
   \int Q_x(x_k) \, Q_{\Sigma}(\Sigma_k) \log\left(
    \frac{Q_x(x_k) \, Q_{\Sigma}(\Sigma_k)}
         {p(x_k,\Sigma_k\,|\,y_{1:k})}
  \right)
  \mathrm{d} x_k \, \mathrm{d} \Sigma_k.
\end{split}
\nonumber
\end{equation}
Minimizing the KL divergence with respect to the probability
densities, we get the following equations:
\begin{equation}
\begin{split}
Q_x(x_k) &\propto
  \exp\left(\int \log p(y_k,x_k,\Sigma_k\,\vert\,y_{1:k-1}) \,
   Q_{\Sigma}(\Sigma_k) \, \mathrm{d} \Sigma_k \right)
  \\
Q_{\Sigma}(\Sigma_k) &\propto
  \exp\left(\int \log p(y_k,x_k,\Sigma_k\,\vert\,y_{1:k-1}) \,
   Q_x(x_k) \, \mathrm{d} x_k \right).
\end{split}
\label{eq:qfp}
\end{equation}
These equations can be interpreted and used as fixed-point iteration
for the sufficient statistics of the approximating densities.

In the original VB-AKF \cite{Sarkka+Nummenmaa:2009}, the VB
approximation was derived for linear state space models with diagonal
noise covariance matrix. In this paper, we generalize it to
non-linear systems with non-diagonal noise covariance matrix.

\section{Main Results}
%

\subsection{Estimation of Full Covariance in Linear Case}
We start by considering the linear state space model with
unknown covariance as follows:
\begin{equation}
\begin{split}
    p(x_k\,|\,x_{k-1}) &=
    \mathrm{N}(x_k\,|\,A_k \, x_{k-1}, Q_k) \\
    p(y_k\,|\,x_{k},\Sigma_k) &=
    \mathrm{N}(x_k\,|\,H_k \, x_k, \Sigma_k),
\end{split}
\label{eq:linmodel}
\end{equation}
where $A_k$ and $H_k$ are some known matrices. We assume that
the dynamic model for the covariance is independent of the state and
of the Markovian form $p(\Sigma_k\,|\,\Sigma_{k-1})$, and set some
restrictions to it shortly. In this section we follow the
derivation in \cite{Sarkka+Nummenmaa:2009}, and extend the scalar
variance case to the full covariance case.

Assume that the filtering distribution of the time step $k-1$
can be approximated as product of Gaussian distribution and
inverse Wishart (IW) distribution as follows:
\begin{equation}
\begin{split}
  &p(x_{k-1},\Sigma_{k-1}\,|\,y_{1:k-1}) = \\
  &\qquad  
  \textrm{N}(x_{k-1} \,\vert\, m_{k-1},P_{k-1}) \,
  \textrm{IW}(\Sigma_{k-1}\,\vert\,\nu_{k-1},V_{k-1}).
\end{split}
\nonumber
\end{equation} 
where the densities, up to non-essential normalization terms, can be
written as \cite{Gelman+Carlin+Stern+Rubin:1995}:
\begin{equation}
\begin{split}
  \textrm{N}(x \,\vert\, m,P)
  &\propto |P|^{-1/2}
      \exp\left( -\frac{1}{2} (x - m)^T \,
        P^{-1} \, (x - m) \right) \\
  \textrm{IW}(\Sigma\,\vert\,\nu,V)   
  &\propto |\Sigma|^{-(\nu+n+1)/2} \, 
     \exp\left( -\frac{1}{2}
      \mathrm{tr}\left( V \, \Sigma^{-1} \right) \right).
\end{split}
\nonumber
\end{equation} 
That is, in the VB approximation \eqref{eq:vbapp}, $Q_x(x_k)$ is the
Gaussian distribution and $Q_{\Sigma}(\Sigma_k)$ is the inverse
Wishart distribution.

We now assume that the dynamic model for the covariance is of
such form that it maps an inverse Wishart distribution at the previous
step into inverse Wishart distribution at the current step. 
That is, the Chapman-Kolmogorov equation \cite{Jazwinski:1970} gives 
the prediction
\begin{equation}
\begin{split}
  &p(\Sigma_k\,\vert\,y_{1:k-1}) \\
  &\qquad = \int p(\Sigma_k\,|\,\Sigma_{k-1}) \, 
  \textrm{IW}(\Sigma_{k-1}\,\vert\,\nu_{k-1},V_{k-1}) \,
  \mathrm{d} \Sigma_{k-1} \\
  &\qquad = \textrm{IW}(\Sigma_{k}\,\vert\,\nu^-_{k},V^-_{k}),
\end{split}
\nonumber
\end{equation} 
for some parameters $\nu^-_{k}$ and $V^-_{k}$. We postpone the
discussion on how to actually calculate these parameters to Section
\ref{sec:dyn}. For the state part we obtain the prediction
\begin{equation}
\begin{split}
  &p(x_k\,\vert\,y_{1:k-1}) \\
  &\qquad = \int \mathrm{N}(x_k\,|\,A_k \, x_{k-1}, Q_k) \,
  \mathrm{N}(x_{k-1}\,|\,m_{k-1}, P_{k-1}) \,
  \mathrm{d} x_{k-1} \\
  &\qquad = \textrm{N}(x_{k}\,\vert\,m^-_{k},P^-_{k}),
\end{split}
\nonumber
\end{equation} 
where $m^-_{k}$ and $P^-_{k}$ are given by the standard Kalman filter
prediction equations:
\begin{equation}
\begin{split}
  m^-_{k} &= A_k \, m_{k-1} \\
  P^-_{k} &= A_k \, P_{k-1} \, A_k^T + Q_k.
\end{split}
\label{eq:kfpred}
\end{equation} 
Because the distribution and the previous step is separable, and the
dynamic models are independent we thus get the following joint
predicted distribution:
\begin{equation}
\begin{split}
  &p(x_{k},\Sigma_{k}\,|\,y_{1:k-1}) = \\
  &\qquad  
  \textrm{N}(x_{k} \,\vert\, m^-_{k},P^-_{k}) \,
  \textrm{IW}(\Sigma_{k}\,\vert\,\nu^-_{k},V^-_{k}).
\end{split}
\nonumber
\end{equation} 
We are now ready to form the actual VB approximation to the posterior.
The integrals in the exponentials of \eqref{eq:qfp}
can now be expanded as follows (cf. \cite{Sarkka+Nummenmaa:2009}):
\begin{equation}
\begin{split}
&\int \log p(y_k,x_k,\Sigma_k\,\vert\,y_{1:k-1}) \,
   Q_{\Sigma}(\Sigma_k) \, \mathrm{d} \Sigma_k \\
&\quad  = -\frac{1}{2} (y_k - H_k\,x_k)^T
   \langle \Sigma_k^{-1} \rangle_{\Sigma}
   (y_k - H_k\,x_k)
  \\ &\qquad
  - \frac{1}{2}(x_k - m^-_k)^T \left(P^-_{k}\right)^{-1} (x_k - m^-_k) + C_1 \\
&\int \log p(y_k,x_k,\Sigma_k\,\vert\,y_{1:k-1}) \,
   Q_{x}(x_k) \, \mathrm{d} x_k \\
&\quad  = -\frac{1}{2} (\nu_k^- + n + 2) \, \log |\Sigma_k|
  - \frac{1}{2} \mathrm{tr}\left\{ V_k^- \, \Sigma_k^{-1} \right\}
  \\ &\qquad 
  - \frac{1}{2} \langle (y_k - H_k\,x_k)^T \,
    \Sigma_k^{-1} \, (y_k - H_k\,x_k) \rangle_x 
  + C_2,
\end{split}
\label{eq:logs}
\end{equation}
where $\langle\cdot\rangle_{\Sigma} = \int (\cdot)
\,Q_{\Sigma}(\Sigma_k) \, \mathrm{d} \Sigma_k$,
$\langle\cdot\rangle_{x} = \int (\cdot) \,Q_{x}(x_k) \, \mathrm{d}
x_k$, and $C_1,C_2$ are some constants. If we have that
$Q_{\Sigma}(\Sigma_k) = \textrm{IW}(\Sigma_k\,\vert\,\nu_k,V_k)$, then
the expectation in the first equation of \eqref{eq:logs} is
\begin{equation}
\begin{split}
  \langle \Sigma_k^{-1} \rangle_{\Sigma}
  &= (\nu_k - n - 1) V_k^{-1}.
\end{split}
\label{eq:exp1}
\end{equation}
Furthermore, if $Q_{x}(x_k) = \textrm{N}(x_{k} \,\vert\,
m_{k},P_{k})$, then the expectation in the second equation of
\eqref{eq:logs} becomes
\begin{equation}
\begin{split}
  &\langle (y_k - H_k\,x_k)^T \,
    \Sigma_k^{-1} \, (y_k - H_k\,x_k) \rangle_x \\
  &\qquad =
    \mathrm{tr}\left\{ 
      H_k \, P_k \, H_k^T \, \Sigma_k^{-1} \right\} \\
  &\qquad + \mathrm{tr}\left\{ (y_k - H_k \, m_k) \,
      (y_k - H_k \, m_k)^T \, \Sigma_k^{-1} \right\}.
\end{split}
\label{eq:exp2}
\end{equation}
By substituting the expectations \eqref{eq:exp1} and \eqref{eq:exp2} into
\eqref{eq:logs} and matching terms in left and right hand sides of
\eqref{eq:qfp} results in the following coupled set of equations:
\begin{equation}
\begin{split}
  S_k &= H_k \, P_k^- \, H_k^T + (\nu_k - n - 1)^{-1} \, V_k \\
  K_k &= P_k^- \, H_k^T \, S_k^{-1} \\
  m_k &= m_k^- + K_k \, (y_k - H_k \, m_k^-) \\
  P_k &= P_k^- - K_k \, S_k \, K_k^T \\
\nu_k &= \nu_k^- + 1 \\
  V_k &= V^-_k + H_k \, P_k \, H_k^T +
   (y_k - H_k \, m_k) \, (y_k - H_k \, m_k)^T.
\end{split}
\label{eq:fpeq}
\end{equation}
The first four of the equations have been written into such suggestive
form that they can easily be recognized to be the Kalman filter update
step equations with measurement noise covariance $(\nu_k - n - 1)^{-1}
\, V_k$.

\subsection{Dynamic Model for Covariance} \label{sec:dyn}
In analogous manner to \cite{Sarkka+Nummenmaa:2009}, the dynamic model
$p(\Sigma_k\,|\,\Sigma_{k-1})$ needs to be chosen such that when it is
applied to an inverse Wishart distribution, it produces another
inverse Wishart distribution. Although, the explicitly construction of
the density is hard, all we need to do is to postulate a
transformation rule for the sufficient statistics of the inverse
Wishart distributions at the prediction step.  Using similar
heuristics as in \cite{Sarkka+Nummenmaa:2009}, we arrive at the
following dynamic model:
\begin{equation}
\begin{split}
  \nu_k^- &= \rho \, (\nu_{k-1} - n - 1) + n + 1 \\
    V_k^- &= B \, V_{k-1} \, B^T,
\end{split}
\end{equation}
where $\rho$ is a real number $0 < \rho \le 1$ and $B$ is a matrix $0
< |B| \le 1$. A reasonable choice for the matrix is $B = \sqrt{\rho}
\, I$, in which case parameter $\rho$ controls the assumed dynamics:
value $\rho = 1$ corresponds to stationary covariance and lower values
allow for higher time-fluctuations. The resulting multidimensional
variational Bayesian adaptive Kalman filter (VB-AKF) is shown in
Algorithm~\ref{alg:linfilter}.
\begin{algorithm}[htb]
\begin{itemize}
\item {\em Predict}: Compute the parameters of the predicted
  distribution as follows:
\begin{equation}
\begin{split}
  m^-_k &= A_k \, m_{k-1} \nonumber\\
  P^-_k &= A_k \, P_{k-1} \, A_{k}^T + Q_k \nonumber\\
  \nu_k^- &= \rho \, (\nu_{k-1} - n - 1) + n + 1 \\
    V_k^- &= B \, V_{k-1} \, B^T,
\end{split}
\nonumber
\end{equation}

\item {\em Update}: First set $m_k^{(0)} = m^-_k$, $P_k^{(0)} =
  P^-_k$, $\nu_{k} = 1 + \nu^-_{k}$, and $V_{k}^{(0)} = V^-_{k}$ and
  the iterate the following, say $N$, steps $i = 1,\ldots,N$:
\begin{equation}
\begin{split}
    S_{k}^{(i+1)} &= H_k \, P_k^- \, H_k^T + (\nu_k - n - 1)^{-1} \, V_k^{(i)} \\
    K_{k}^{(i+1)} &= P_k^- \, H_k^T \, [S_{k}^{(i+1)}]^{-1} \\
    m_{k}^{(i+1)} &= m^-_{k} + K_{k}^{(i+1)} \, (y_k - H_k \, m_k) \\
    P_{k}^{(i+1)} &= P^-_{k} - K_{k}^{(i+1)} \,
      S_{k}^{(i+1)} \, [K_{k}^{(i+1)}]^T \\
      V_k^{(i+1)} &= V^-_k + H_k \, P_k^{(i)} \, H_k^T +
       (y_k - H_k \, m_k^{(i)}) \, (y_k - H_k \, m_k^{(i)})^T
\end{split}
\nonumber
\end{equation}
and set $V_{k} = V_{k}^{(N)}$, $m_k = m_k^{(N)}$, $P_k =
P_k^{(N)}$.
\end{itemize}
\caption{The multidimensional Variational Bayesian Adaptive Kalman Filter (VB-AKF)
  algorithm}
\label{alg:linfilter}
\end{algorithm}

\subsection{Extension to Non-Linear Models}
In this section we extend the results in the previous section
into non-linear models of the form \eqref{eq:ssmodel1}. We start with
the assumption that the filtering distribution is approximately
product of a Gaussian term and inverse Wishart (IW) term:
\begin{equation}
\begin{split}
  &p(x_{k-1},\Sigma_{k-1}\,|\,y_{1:k-1}) = \\
  &\qquad  
  \textrm{N}(x_{k-1} \,\vert\, m_{k-1},P_{k-1}) \,
  \textrm{IW}(\Sigma_{k-1}\,\vert\,\nu_{k-1},V_{k-1}).
\end{split}
\nonumber
\end{equation} 
The prediction step can be handled in similar manner as in the linear
case, except that the computation of the mean and covariance of the
state should be done with the Gaussian filter prediction equations
\eqref{eq:adf_predict1} instead of the Kalman filter prediction
equations \eqref{eq:kfpred}. The inverse Wishart part of the prediction
remains intact.

After the prediction step, the approximation is then
\begin{equation}
\begin{split}
  &p(x_{k},\Sigma_{k}\,|\,y_{1:k-1}) = \\
  &\qquad  
  \textrm{N}(x_{k} \,\vert\, m^-_{k},P^-_{k}) \,
  \textrm{IW}(\Sigma_{k}\,\vert\,\nu^-_{k},V^-_{k}).
\end{split}
\nonumber
\end{equation} 
The expressions corresponding to \eqref{eq:logs} now become:
\begin{equation}
\begin{split}
&\int \log p(y_k,x_k,\Sigma_k\,\vert\,y_{1:k-1}) \,
   Q_{\Sigma}(\Sigma_k) \, \mathrm{d} \Sigma_k \\
&\quad  = -\frac{1}{2} (y_k - h(x_k))^T
   \langle \Sigma_k^{-1} \rangle_{\Sigma}
   (y_k - h(x_k))
  \\ &\qquad
  - \frac{1}{2}(x_k - m^-_k)^T \left(P^-_{k}\right)^{-1} (x_k - m^-_k) + C_1 \\
&\int \log p(y_k,x_k,\Sigma_k\,\vert\,y_{1:k-1}) \,
   Q_{x}(x_k) \, \mathrm{d} x_k \\
&\quad  = -\frac{1}{2} (\nu_k^- + n + 2) \, \log |\Sigma_k|
  - \frac{1}{2} \mathrm{tr}\left\{ V_k^- \, \Sigma_k^{-1} \right\}
  \\ &\qquad 
  - \frac{1}{2} \langle (y_k - h(x_k))^T \,
    \Sigma_k^{-1} \, (y_k - h(x_k)) \rangle_x 
  + C_2.
\end{split}
\label{eq:logs2}
\end{equation}
The expectation in the first equation is still given by the equation
\eqref{eq:exp1}, but the resulting distribution in terms of $x_k$ is
intractable in closed form due to the non-linearity $h(x_k)$.
Fortunately, the approximation problem is exactly the same as
encountered in the update step of Gaussian filter and thus we can
directly use the equations \eqref{eq:adf_update1} for computing
Gaussian approximation to the distribution.

The simplification \eqref{eq:exp2} does not work in the non-linear
case, but we can rewrite the expectation as
\begin{equation}
\begin{split}
  &\langle (y_k - h(x_k))^T \,
    \Sigma_k^{-1} \, (y_k - h(x_k)) \rangle_x \\
  &\qquad =
    \mathrm{tr}\left\{    
    \langle (y_k - h(x_k)) \, (y_k - h(x_k))^T \rangle_x \, \Sigma_k^{-1}
    \right\},
\end{split}
\label{eq:exp3}
\end{equation}
where the expectation can be separately computed using some of the
Gaussian integration methods in \cite{Wu+Hu+Wu+Hu:2006}.  Because the
result of the integration is just a constant matrix, we can now
substitute \eqref{eq:exp1} and \eqref{eq:exp3} into \eqref{eq:logs2}
and match the terms in equations \eqref{eq:qfp} in the same manner as
in linear case to obtain the equations:
\begin{equation}
\begin{split}
    \mu_{k} &= 
       \int h(x_{k}) \,
        \mathrm{N}(x_{k}\,|\,m^-_{k},P^-_{k}) \, \mathrm{d} x_k \\
    S_{k} &= \int (h(x_{k}) - \mu_{k}) \,
       (h(x_{k}) - \mu_{k})^T \\
     &\qquad \times 
       \mathrm{N}(x_{k}\,|\,m^-_{k},P^-_{k}) \, \mathrm{d} x_{k}
       + (\nu_k - n - 1)^{-1} \, V_k \\
    C_k &=  \int (x_k - m^-) \,
                  (h(x_k) - \mu_k)^T \,
                  \mathrm{N}(x_k\,|\,m^-_k,P^-_k) \, \mathrm{d} x_k \\
    K_{k} &= C_k \, S^{-1}_{k} \\
    m_{k} &= m^-_{k} + K_{k} \, (y_k - \mu_k) \\
    P_{k} &= P^-_{k} - K_{k} \, S_{k} \, K^T_{k} \\
 \nu_k &= \nu_k^- + 1 \\
   V_k &= V^-_k + 
    \int (y_k - h(x_k)) \, (y_k - h(x_k))^T \\
   &\qquad \qquad \times \mathrm{N}(x_k\,|\,m_k,P_k) \, \mathrm{d} x_k.
\end{split}
\label{eq:fpeq2}
\end{equation}

\subsection{The Adaptive Filtering Algorithm}
\begin{algorithm}[htb]
\begin{itemize}
\item {\em Predict}: Compute the parameters of the predicted
  distribution as follows:
\begin{equation}
\begin{split}
    m^-_{k} &=
      \int f(x_{k-1}) \,
        \mathrm{N}(x_{k-1}\,|\,m_{k-1},P_{k-1}) \, \mathrm{d} x_{k-1} \\
    P^-_{k} &= \int (f(x_{k-1}) - m^-_{k}) \,
       (f(x_{k-1}) - m^-_{k})^T \\
       &\qquad \times
       \mathrm{N}(x_{k-1}\,|\,m_{k-1},P_{k-1}) \, \mathrm{d} x_{k-1} 
       + Q_{k} \\
  \nu_k^- &= \rho \, (\nu_{k-1} - n - 1) + n + 1 \\
    V_k^- &= B \, V_{k-1} \, B^T,
\end{split}
\nonumber
\end{equation}

\item {\em Update}: First set $m_k^{(0)} = m^-_k$, $P_k^{(0)} =
  P^-_k$, $\nu_{k} = 1 + \nu^-_{k}$, and $V_{k}^{(0)} = V^-_{k}$
  and precompute the following:
\begin{equation}
\begin{split}
    \mu_{k} &= 
       \int h(x_{k}) \,
        \mathrm{N}(x_{k}\,|\,m^-_{k},P^-_{k}) \, \mathrm{d} x_k \\
    T_{k} &= \int (h(x_{k}) - \mu_{k}) \,
       (h(x_{k}) - \mu_{k})^T \\
     &\qquad \times 
       \mathrm{N}(x_{k}\,|\,m^-_{k},P^-_{k}) \, \mathrm{d} x_{k} \\
    C_k &=  \int (x_k - m^-) \,
                 (h(x_k) - \mu_k)^T \\
    &\qquad \times \mathrm{N}(x_k\,|\,m^-_k,P^-_k) \, \mathrm{d} x_k \\
\end{split}
\nonumber
\end{equation}
Iterate the following, say $N$, steps $i = 1,\ldots,N$:
\begin{equation}
\begin{split}
    S_{k}^{(i+1)} &= T_{k} + (\nu_k - n - 1)^{-1} \, V_k^{(i)} \\
    K_{k}^{(i+1)} &= C_k \, [S_{k}^{(i+1)}]^{-1} \\
    m_{k}^{(i+1)} &= m^-_{k} + K_{k}^{(i+1)} \, (y_k - \mu_k) \\
    P_{k}^{(i+1)} &= P^-_{k} - K_{k}^{(i+1)} \,
      S_{k}^{(i+1)} \, [K_{k}^{(i+1)}]^T \\
      V_k^{(i+1)} &= V^-_k 
    + \int (y_k - h(x_k)) \, (y_k - h(x_k))^T \\
    &\qquad \qquad \times \mathrm{N}(x_k\,|\,m_k^{(i)},P_k^{(i)}) \, \mathrm{d} x_k. \\
\end{split}
\nonumber
\end{equation}
and set $V_{k} = V_{k}^{(N)}$, $m_k = m_k^{(N)}$, $P_k =
P_k^{(N)}$.
\end{itemize}
\caption{The Variational Bayesian Adaptive Gaussian Filter (VB-AGF)
  algorithm}
\label{alg:filter}
\end{algorithm}
The general filtering method for the full covariance and non-linear
state space model is shown in Algorithm~\ref{alg:filter}. Various
useful special cases and extensions can be deduced from the equations:
\begin{itemize}
\item The {\em Gaussian integration method} will result in different
  variants of the algorithm. For example, the Taylor series based
  approximation could be called VB-AEKF, unscented transform based
  method VB-AUKF, cubature based VB-ACKF, Gauss-Hermite based VB-AGHKF
  and so on. For the details of the different Gaussian integration
  methods, see,
  \cite{Ito+Xiong:2000,Wu+Hu+Wu+Hu:2006,Julier+Uhlmann+Durrant-Whyte:2000,Arasaratnam+Haykin:2009,Kotecha+Djuric:2003,Norgaard+Poulsen+Ravn:2000,Lefebvre+Bruyninckx+DeSchutter:2002,Deisenroth+Huber+Hanebeck:2009}.

\item The {\em linear special case} can be easily deduced by comparing
  the equations \eqref{eq:fpeq} and \eqref{eq:fpeq2} to the equations in the
  algorithm. That is, one simply replaces the Gaussian integrals with their
  closed form solutions.

\item The {\em diagonal covariance case}, which was considered in
  \cite{Sarkka+Nummenmaa:2009}, can be recovered by updating only the
  diagonal elements in the last equation of the Algorithm and keeping
  all other elements in the matrices $V_k^{(i)}$ zero. Of course, the
  matrix $B$ in the prediction step then needs to be diagonal also.
  Note that the inverse Wishart parameterization does not reduce to the inverse
  Gamma parameterization, but still the formulations are equivalent.

\item {\em Non-additive dynamic models} can be handled by simply
  replacing the state prediction with the non-additive counterpart.
\end{itemize}


%
%
%
%
%
%
%
%
%
%
%
%
%
%
%
%
%
%







\section{Numerical Results}
\subsection{Range-Only Tracking in a Non-homogeneous Noise Field}
In this simple example we illustrate the performance of the developed
adaptive filters by tracking a moving target with sensors, which
measure the distances to the target moving in 2-dimensional $(u,v)$
space. The measurements are corrupted with noise having time-varying
correlations between the sensors. The correlations arise, because the
noise in the measurements is caused by localized variations in the
environment and when the spatial paths of the measured signals are similar,
the noises are correlated.

The state is contains the position and velocity of the target
$x_k = (u_k\,v_k\;\dot{u}_k\;\dot{v}_k)^T$ and the dynamics
of the target are modeled by the standard Wiener velocity model.
%
%
%
%
The distance measurements from $m$ sensors read
\begin{equation}
  y_k^i = \sqrt{(s_u^i - u_k)^2 + (s_v^i - v_k)^2}
  + r_k^i, \;\;\; i = 1,\ldots,m,
\end{equation}
where $(s_u^i, s_v^i)$ is the position of $i$th sensor and $r_k^i$ is
the $i$th component of a Gaussian distributed noise vector $r_k
\sim N(0,\Sigma_k)$. In this experiment the noise to each
distance measurement is generated by drawing a random sample from a
discretized Gaussian random field $z_k \in \mathbb{R}^{n_z}$ and
then collecting all the values of the field connected to the line
between the sensor and the target.
We take the time-white
continuous-valued random field $z(u,v)$ to be zero mean and to
have the covariance function
\begin{equation}
\begin{split}
  k(u,v,u',v') & = \sigma_{\text{bg}}^2 \delta(u-u') \delta(v-v') \\
& + \sum_{i=1}^{n_A}  1_{A_i}(u,v) \, 1_{A_i}(u',v') \, k_i(u,v,u',v'),
\end{split}
\end{equation}
where the first term corresponds to white background noise and the
latter terms to additive correlations for points inside bounded regions
$A_i \subset \mathbb{R}^2 ,i=1\,\ldots,n_A$ with covariance functions
$k_i(u,v,u',v')$. We set the covariance functions to
\begin{equation}
\begin{split}
k_i(u,v,u',v') & = \sigma_{\text{bg},i}^2 \, \delta(u-u') \, \delta(v-v') \\
& + \sigma_{\text{magn},i}^2 \exp\left(\frac{1}{l_i^2}((u-u')^2+(v-v')^2) \right),
\end{split}
\end{equation}
which means that noise inside the bounded regions consists of
independent (white) and correlated components.

\begin{figure}
  \begin{center}
    \includegraphics[width=6cm]{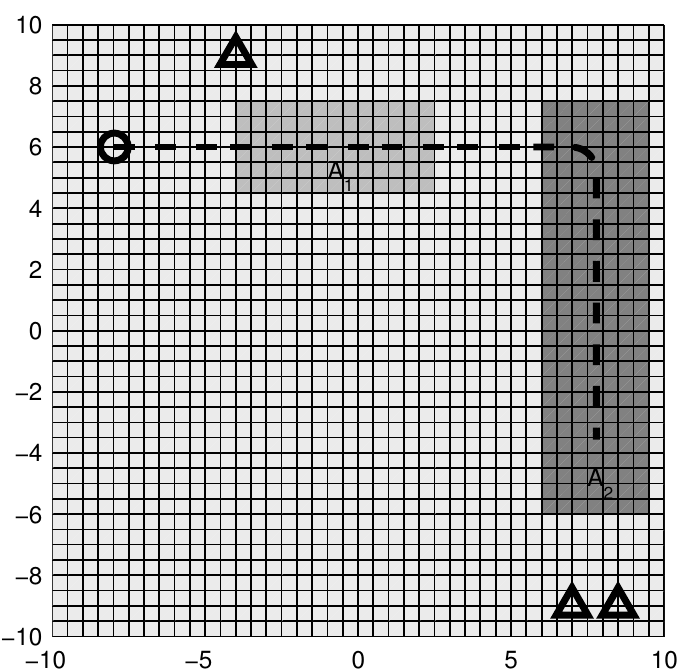}
    \caption{Simulation scenario in the Range-Only Tracking
      Demonstration. Circle denotes the starting location of the
      target, triangles the locations of the sensors and the dashed
      line the trajectory of the target. Inside the shaded areas the
      noise field has spatial correlations. }
    \label{fig:experiment_setup}
  \end{center}
\end{figure}

\begin{figure}
  \begin{center}
    \includegraphics[width=8cm]{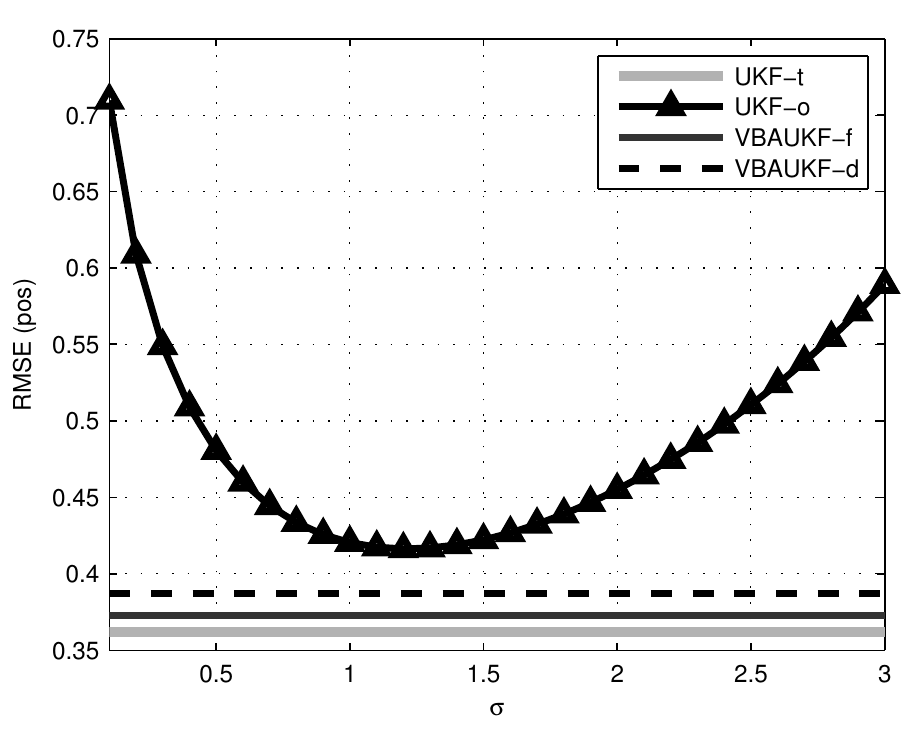}
    \caption{Range-Only Tracking: RMSE values for tracking the
      position of the target on a typical simulation run. The different
      standard deviation values used for UKF-o are on X-axis.}
    \label{fig:position_rmse}
  \end{center}
\end{figure}

\begin{figure}
  \begin{center}
\subfigure[$\sigma^2_{1,1}$]{\includegraphics[width=2.85cm]{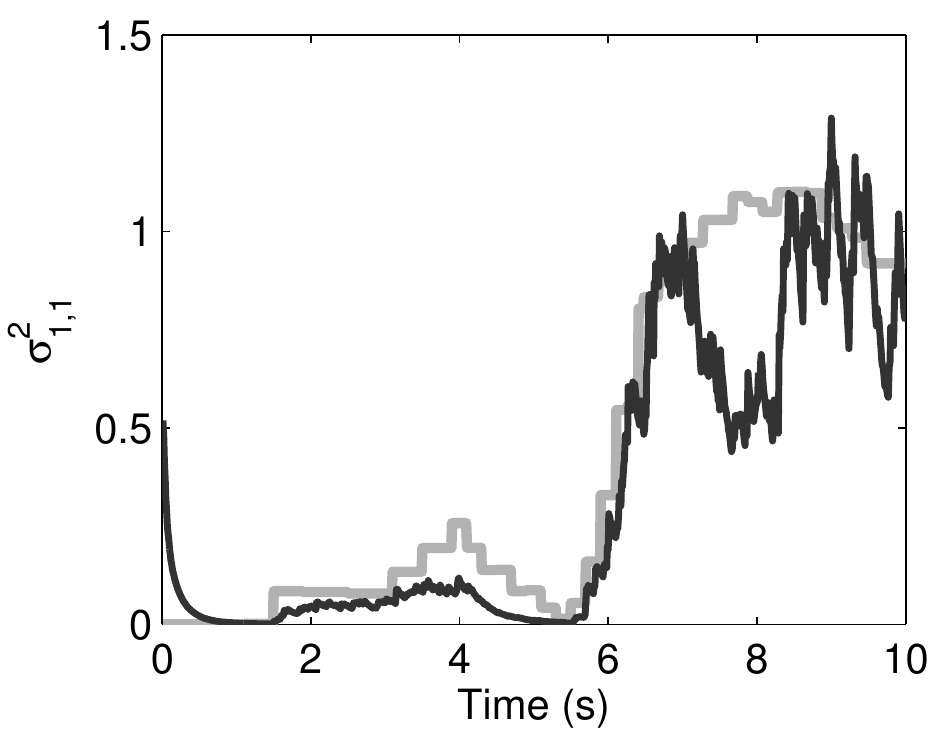}}
\subfigure[$\sigma^2_{2,2}$]{\includegraphics[width=2.85cm]{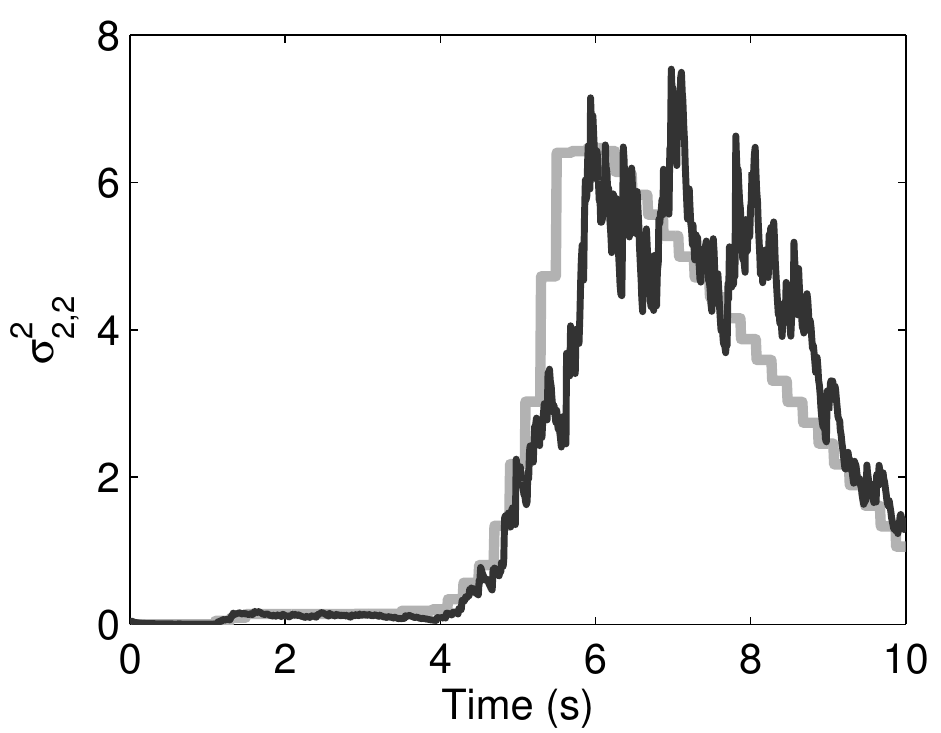}}
\subfigure[$\sigma^2_{3,3}$]{\includegraphics[width=2.85cm]{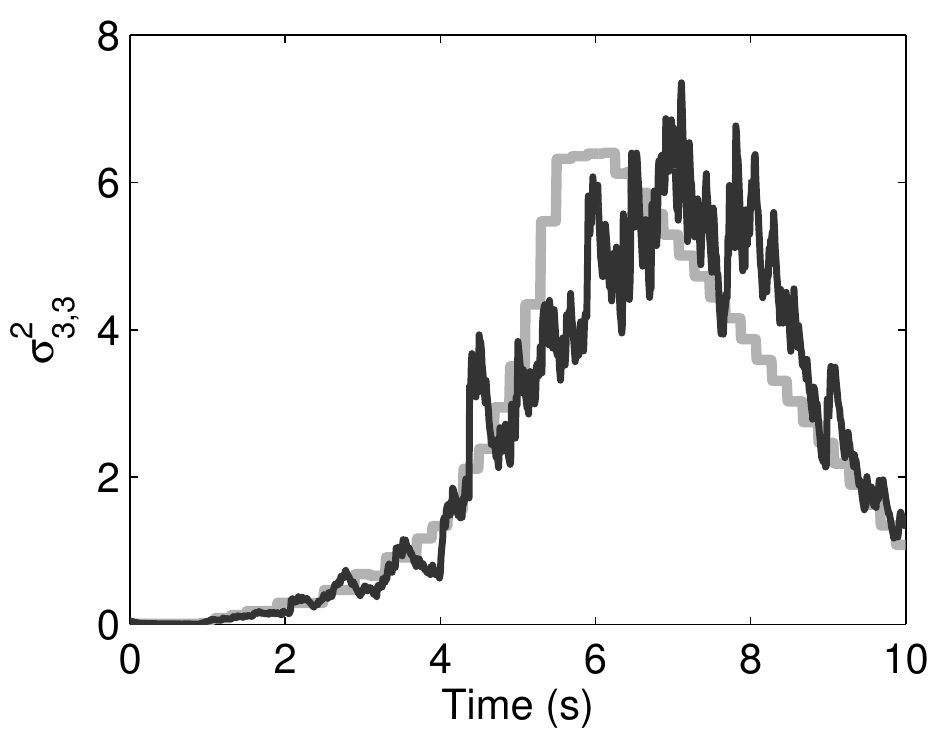}}
\subfigure[$\sigma^2_{1,2}$]{\includegraphics[width=2.85cm]{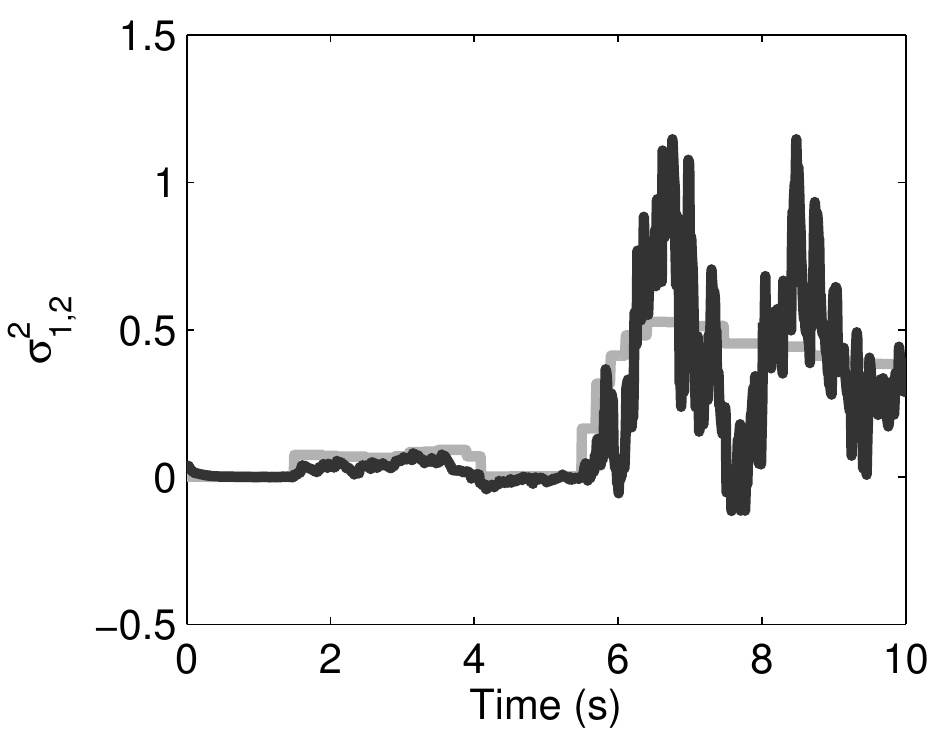}}
\subfigure[$\sigma^2_{1,3}$]{\includegraphics[width=2.85cm]{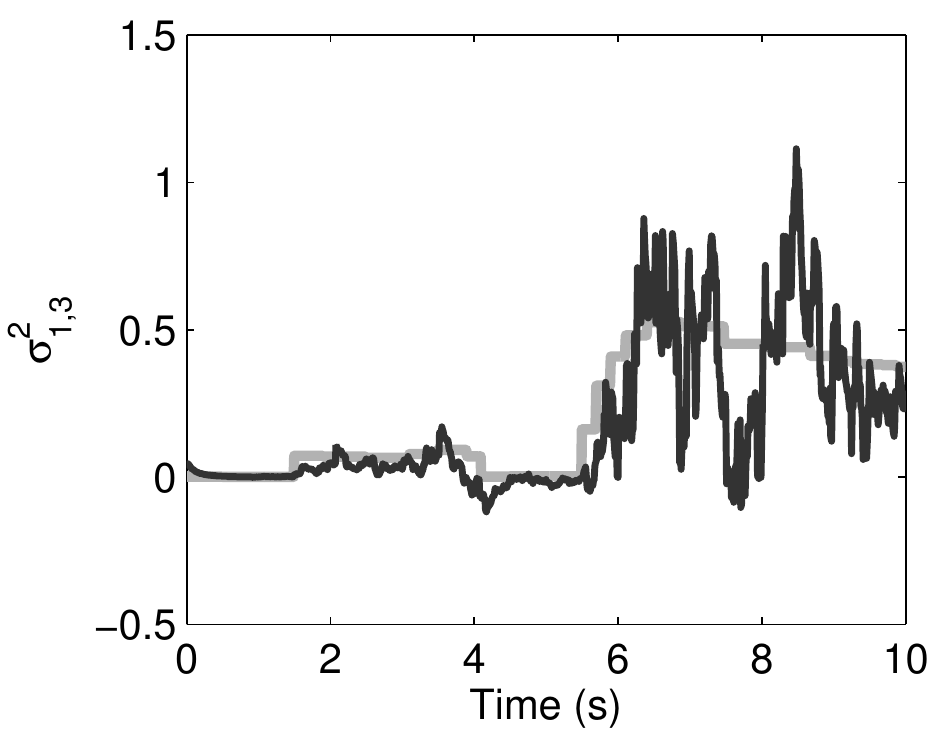}}
\subfigure[$\sigma^2_{2,3}$]{\includegraphics[width=2.85cm]{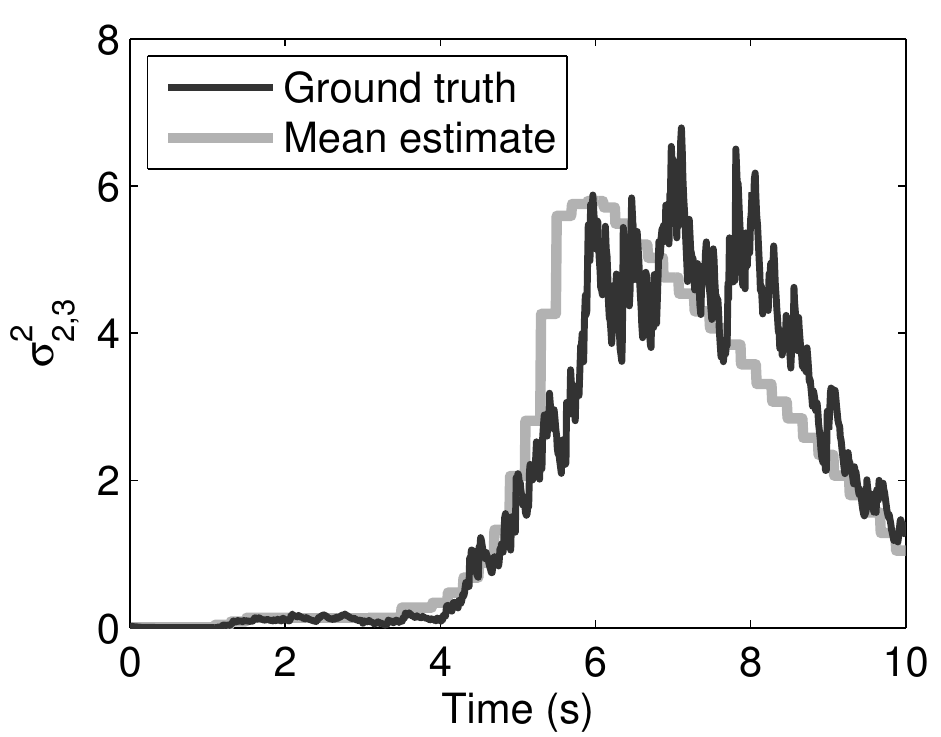}}
\caption{Range-Only Tracking: Estimates of $\Sigma_k$ with VBAUKF-f.
  Smoother estimates can be obtained by tuning $\rho$ to be larger,
  but that introduces more lag to estimates.}
    \label{fig:covariance_estimate}
  \end{center}
\end{figure}

The simulation scenario is illustrated in Figure
\ref{fig:experiment_setup}. For the lightly shaded area ($A_1$) the
covariance function parameters were $\sigma_{\text{bg},i}^2 = 0.01^2$,
$\sigma_{\text{magn},i}^2 = 0.1^2$ and $l_i = 2$, and inside the
darkly shaded area ($A_2$) $\sigma_{\text{bg},i}^2 = 0.01^2$,
$\sigma_{\text{magn},i}^2 = 0.2^2$ and $l_i = 2$. The variance of the
background noise was set to $\sigma_{\text{bg}}^2 = 0.01^2$. The
spectral density of the process noise was set to $q=2$ and the time
step to $T = 0.01$. The trajectory shown in Figure
\ref{fig:experiment_setup} was discretized to $1000$ time steps and
then measurements were generated according to the procedure described
above. Given the measurements, the target was tracked with the
following methods:
\begin{itemize}
\item UKF-t: Unscented Kalman filter with measurement covariance set
  to true value on each time step.
\item UKF-o: Unscented Kalman filter with fixed diagonal measurement
  covariance matrix with different standard deviations $\sigma =
  0.1,\ldots, 3$.
\item VB-AUKF-f: The proposed adaptive filter with ADF approximations
  made with UKF. The parameter $\rho$ in dynamic model of measurement
  noise was set to $\rho = 1-\exp(-3)$.
\item VB-AUKF-d: Same as VB-AUKF-f with the exception that the
  measurement covariance is forced to be diagonal.
\end{itemize}
With all methods the parameters of UKF was set to default values
$\alpha = 1,\beta = 0,\kappa = 3-m$. The RMSE values in tracking the
position of the target with the tested methods are shown in Figure
\ref{fig:position_rmse} for a typical simulation run. It can be seen
that the best results can be achieved with exact measurement covariance
while the estimation of full covariance improves the results of VB-AUKF
over the diagonal case. Obviously, UKF with fixed diagonal measurement
covariance is clearly the worst of the all the tested methods. Figure
\ref{fig:covariance_estimate} shows the estimates of the element of
$\Sigma_k$ produced by VB-AUKF-f together with their true values.

\subsection{Multi-Sensor Bearings Only Tracking}
As an example, we consider the classical multi-sensor bearings only
tracking problem with coordinated turning model, where the state $x =
(u,\dot{u},v,\dot{v},\omega)$ contains the 2d location $(u,v)$ and the
corresponding velocities $(\dot{u},\dot{v})$ as well as the turning
rate $\omega$ of the target. The dynamic model and the measurement
models for sensors $i=1,\ldots,4$ are given as:
\begin{equation}
\begin{split}
x_k &= \begin{pmatrix} 1 & \frac{\sin(\omega \Delta t)}{\omega} & 0 & -
  \left( \frac{1-\cos(\omega \Delta t)}{\omega} \right)& 0 \\
0 & \cos(\omega \Delta t) & 0 & -\sin(\omega \Delta t) & 0 \\
0 & \frac{1-\cos(\omega \Delta t)}{\omega }& 1 & \frac{\sin(\omega
  \Delta t)}{\omega} & 0 \\
0 & \sin(\omega \Delta t) & 0 & \cos(\omega \Delta t) & 0 \\
0 & 0 & 0 & 0 & 1 \end{pmatrix} x_{k-1} + q_{k-1} \\
    y_{k} &= \arctan\left(
    \frac{v_{k} - s^i_v}{u_{k} - s^i_u} \right) + r_{i,k}, \qquad i=1,\ldots,4,
\end{split}
\end{equation}
where $q_k \sim \mathrm{N}(0,Q)$ is the process noise and $r_k =
(r_{k,1},\ldots,r_{k,4})$ are the measurement noises of sensors with
joint distribution $r_k \sim \mathrm{N}(0,\Sigma_k)$, where $\Sigma_k$
is unknown and time varying.

\begin{figure}[!t]
\centering
\includegraphics[width=0.45\textwidth]{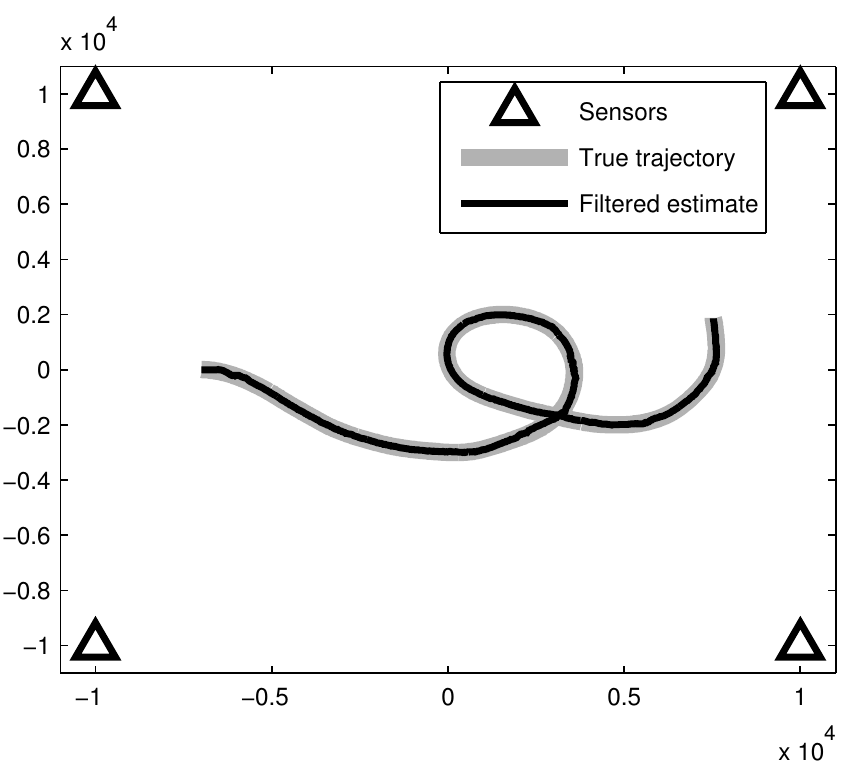}
\caption{The simulated trajectory and the estimate obtained
with VB-ACKF.}
\label{fig:traj}
\end{figure}

\begin{figure}[!t]
\centering
\includegraphics[width=0.45\textwidth]{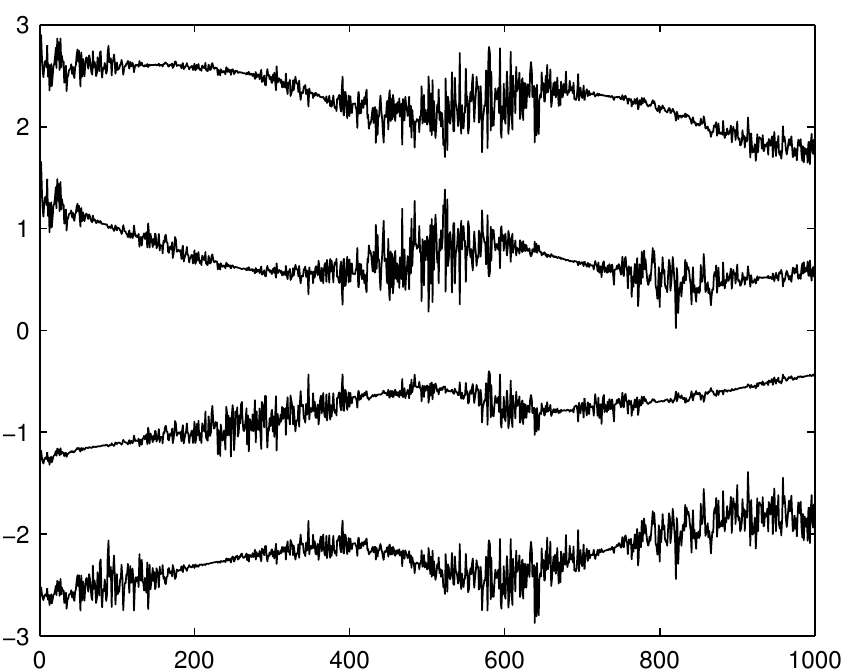}
\caption{The simulated measurements.}
\label{fig:meas}
\end{figure}

We simulated a trajectory and measurements from the model and applied
different filters to it. We tested various Gaussian integration based
methods (VB-AEKF, VB-AUKF, VB-ACKF, VB-AGHKF) and because the results
were quite much the same with different Gaussian integration methods
(though VB-AEKF was a bit worse than the others), we only present the
results obtained with VB-ACKF.  Figure~\ref{fig:traj} shows the
simulated trajectory and the VB-ACKF results with the full covariance
estimation.  In the simulation, the variances of the measurement
noises as well as the cross-correlations varied smoothly over time.
The simulated measurements are shown in Figure~\ref{fig:meas}. 

\begin{figure}[!t]
\centering
\includegraphics[width=0.45\textwidth]{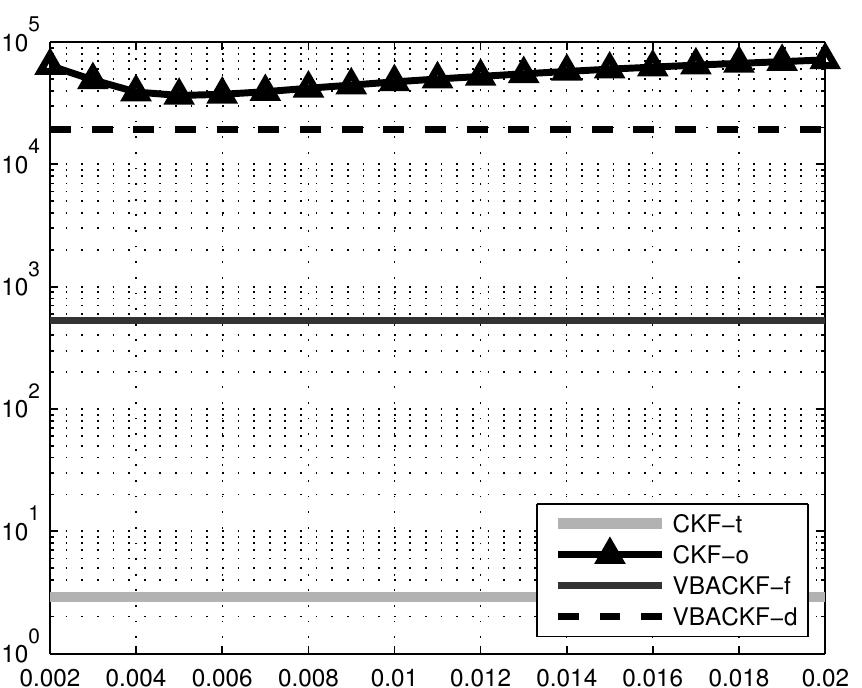}
\caption{Root mean squared errors (RMSEs) for different methods.}
\label{fig:rmse}
\end{figure}

Figure~\ref{fig:rmse} shows the root mean squared errors (RMSEs) for the
following methods:
\begin{itemize}
\item {\em CKF-t}: CKF with the true covariance matrix.
\item {\em CKF-o}: CKF with a diagonal covariance matrix with diagonal
  elements given by the value on the $x$-axis.
\item {\em VBCKF-f}: CKF with full covariance estimation.
\item {\em VBCKF-d:} CKF with diagonal covariance estimation.
\end{itemize}
As can be seen from the figure, the results of filters with covariance
estimation are indeed better than the results of any filter with fixed
diagonal covariance matrix. The filter with the known covariance
matrix gives the lowest error, as would be expected, and the filter
with full covariance estimation gives a lower error than the filter
with diagonal covariance estimation.

\section{Conclusion and Discussion}
In this paper, we have presented a variational Bayes and Gaussian
filtering based algorithm for joint estimation of state and
time-varying noise covariances in non-linear state space models. The
performance of the method has been illustrated in simulated
applications.

There are several extensions that could be considered as well. For
instance, we could try to estimate the process noise covariance in the
model. However, it is not as easy as it sounds, because the process
noise covariance does not appear in the equations in such simple
conjugate form as the measurement noise covariance. Another natural
extension would be the case of smoothing (cf.
\cite{Piche+Sarkka+Hartikainen:2012}). Unfortunately the current
dynamic model makes things challening, because we do not know the
actual transition density at all. This makes the implementation of a
Rauch--Tung--Striebel type of smoother impossible---although a simple
smoothing estimate for the state can be obtained by simply running the
RTS smoother over the state estimates while ignoring the noise
covariance estimates completely. However, it would be possible to
construct an approximate two-filter smoother for the full state space
model, but even in that case we need to put some more constraints to
the model, for example, assume that the covariance dynamics are
time-reversible.





\bibliographystyle{elsarticle-num}
\bibliography{mvb-akf}







\end{document}